\documentclass[prb,aps,epsf,twocolumn,showpacs,superscriptaddress]{revtex4}
\usepackage[dvips]{graphicx}
\usepackage{latexsym}
\usepackage{amsmath}
\begin{document}
\newcommand{\beq}{\begin{equation}}
\newcommand{\beqr}{\begin{eqnarray}}
\newcommand{\eeqr}{\end{eqnarray}}
\newcommand{\eeq}{\end{equation}}
\newcommand{\s}{{\sigma}}
\newcommand{\e}{{\varepsilon}}
\newcommand{\om}{{\omega}}
\newcommand{\Om}{{\Omega}}
\newcommand{\D}{{\Delta}}
\newcommand{\de}{{\delta}}
\newcommand{\al}{{\alpha}}
\newcommand{\ga}{{\gamma}}
\newcommand{\La}{{\Lambda}}
\newcommand{\be}{{\beta}}
\newcommand{\psib}{{\bar{\psi}}}
\newcommand{\rb}{{\bar{\rho}}}
\newcommand{\phib}{{\bar{\phi}}}
\newcommand{\dt}{{\Delta}}
\newcommand{\dva}{{\frac{\vp\times\va}{2\pi}-\rb}}
\newcommand{\dvab}{{\frac{\vp\times(\va-\vb)}{4p\pi}}}
\newcommand{\w}{{\omega}}
\newcommand{\zh}{{\hat{z}}}
\newcommand{\qh}{{\hat{q}}}
\newcommand{\vA}{{\vec{A}}}
\newcommand{\va}{{\vec{a}}}
\newcommand{\vrr}{{\vec{r}}}
\newcommand{\vj}{{\vec{j}}}
\newcommand{\vE}{{\vec{E}}}
\newcommand{\vB}{{\vec{B}}}
\def\bA{{\mathbf A}}
\def\bm{{\mathbf m}}
\def\bsig{{\mathbf \sigma}}
\def\bB{{\mathbf B}}
\def\bp{{\mathbf p}}
\def\bI{{\mathbf I}}
\def\bn{{\mathbf n}}
\def\bM{{\mathbf M}}
\def\bq{{\mathbf q}}
\def\br{{\mathbf r}}
\def\bs{{\mathbf s}}
\def\bS{{\mathbf S}}
\def\bQ{{\mathbf Q}}
\def\bs{{\mathbf s}}
\def\bB{{\mathbf B}}
\def\bl{{\mathbf l}}
\def\bPi{{\mathbf \Pi}}
\def\bJ{{\mathbf J}}
\def\bR{{\mathbf R}}
\def\bz{{\mathbf z}}
\def\ba{{\mathbf a}}
\def\bk{{\mathbf k}}
\def\bK{{\mathbf K}}
\def\bP{{\mathbf P}}
\def\bg{{\mathbf g}}
\def\bX{{\mathbf X}}
\newcommand{\etab}{\mbox{\boldmath $\eta $}}
\newcommand{\sigmab}{\mbox{\boldmath $\sigma $}}
\newcommand{\vx}{{\vec{x}}}
\newcommand{\vq}{{\vec{q}}}
\newcommand{\vQ}{{\vec{Q}}}
\newcommand{\vd}{{\vec{d}}}
\newcommand{\vb}{{\vec{b}}}
\newcommand{\vp}{{\vec{\partial}}}
\newcommand{\p}{{\partial}}
\newcommand{\gr}{{\nabla}}
\newcommand{\ra}{{\rightarrow}}
\def\dd{d^{\dagger}}
\def\half{{1\over2}}
\def\third{{1\over3}}
\def\twof{{2\over5}}
\def\threes{{3\over7}}
\def\rhob{{\bar \rho}}
\def\ua{\uparrow}
\def\da{\downarrow}
\def\eqa{\begin{eqnarray}}
\def\eea{\end{eqnarray}}
\parindent=4mm
\addtolength{\textheight}{0.9truecm}
\title{Fermi surfaces in general co-dimension and a new controlled non-trivial fixed point}
\author{T. Senthil}
\affiliation{ Department of Physics, Massachusetts Institute of
Technology, Cambridge, Massachusetts 02139}
\author{R. Shankar}
\affiliation{Department of Physics, Yale University, New Haven CT 06520}
\date{\today}

\begin{abstract}
Traditionally  Fermi surfaces for problems in $d$ spatial dimensions have dimensionality $d-1$,  i.e., codimension $d_c=1$ along which energy varies.  Situations with $d_c >1$ arise when the gapless fermionic excitations live at isolated nodal points or lines. For $d_c > 1$  weak short range interactions are irrelevant at the non-interacting fixed point.
Increasing interaction strength can lead to phase transitions out of this Fermi liquid. We illustrate this  by studying the  transition to superconductivity  in a controlled $\epsilon$ expansion near $d_c = 1$.
The resulting non-trivial fixed point is shown to describe a scale invariant theory that lives in effective space-time dimension $D=d_c + 1$. Remarkably, the results can be reproduced by the more familiar Hertz-Millis action for the bosonic superconducting order parameter even though it lives in different space-time dimensions.
\end{abstract}

\maketitle

The ground state of a fluid of non-interacting fermions in spatial dimension $d \geq 2$ is typically a Landau fermi liquid. This state has a sharp Fermi surface and well defined quasiparticle excitations. The Fermi liquid state can be understood within a fermionic renormalization group framework\cite{rg,rsrg} in terms of a fixed point obtained as one focuses on modes within a bandwidth $\La$ of the Fermi surface and systematically reduces $\La$. Phase transitions out of the Fermi liquid phase (such as the Stoner ferromagnetic transition) are not described by traditional flows  within this approach. An alternate approach of   Hertz \cite{hertz}, extended by Millis \cite{millis}, performs an  RG analysis on the bosonic action for the order parameter obtained by  integrating out the fermions. This bosonic action  is nonanalytic in frequency and momentum and  yields results  in agreement with those of Moriya \cite{moriya}  who used a self-consistent renormalization approach. Since the approach of integrating out the gapless fermionic modes  is contrary to the dogma of the RG that singularities should appear, not in the action but in the Green's functions, it  has been a source of concern to some.

In situations where the non-interacting band structure consists of distinct Fermi points (such as in graphene) the situation is much better. The low energy theory of such fermion systems is a massless `relativistic' Dirac theory. In dimension $d \geq 2$ weak short ranged interactions are irrelevant in such a theory. At finite interaction strength
various phase transitions that gap out the massless fermions can occur. Without the complications of an extended Fermi surface these can be analysed within the conventional field theoretic framework for critical phenomena.

In this paper we consider a general class of problems that fall in between these two cases. The low energy theory of a system of non-interacting fermions in $d$ spatial dimension may be characterized by the co-dimension of the surface in momentum space where the energy gap vanishes. The ordinary Fermi surface has co-dimension $1$ while the case of Fermi points has co-dimension $d$. It is interesting to study the general case of co-dimension $d_c$ with a eye toward gaining insight into the all important case $d_c = 1$. In dimension $d =  3$ the case $d_c = 2$ is the one new possibility. This corresponds to fermions with line nodes in three dimensions which arise in a number of different situations, most strikingly in several unconventional three dimensional superconductors.

We first use the fermionic RG methods of Ref. \onlinecite{rg,rsrg} to argue that whenever $d_c > 1$ all short range interactions are irrelevant. With increasing interaction strength phase transitions which gap out the fermions are
again possible. For $d_c < d$ these occur in the presence of an extended gapless `Fermi surface' and thus share the complications of the usual case $d_c = 1$. As a concrete example we study a pairing phase transition associated with superconductivity. We show that this can be accessed within the fermionic RG through a controlled $\epsilon$ expansion in the co-dimension $d_c$ near $d_c =1$. Remarkably a non-trivial scaling structure with effective space-time dimensionality $d_c + 1$ is obtained. The transition can also be analysed within the Moriya-Hertz-Millis approach which works with a bosonic order parameter that lives in the full $d$ space dimensions. Despite the difference in dimensionalities both approaches give identical results! Comparison of the two approaches thus provides much insight into the nature of quantum criticality in fermionic systems
with an extended gapless Fermi surface.

Consider then a `generalized Fermi surface' of dimension $d-d_c$. Coordinates on this Fermi surface play the role of  internal symmetry degrees of freedom and do not play a big role in the low energy limit for weak interaction strengths. We refer to them as isospin directions. We will assume a dispersion where the energy varies linearly with momentum deviation away from this surface (as happens for usual Fermi surfaces and for line nodes in $3d$).
Let us call these directions along which energy varies linearly with momentum the Dirac directions.

We begin in $d=2$ when we have a Fermi line  (one isospin direction, parameterized by the angle $\theta$) and one radial Dirac direction measured by $k$.  In this world the {\em  momentum transfer $\bq$} is two-dimensional. We are interested in $\bq$ because  any bosonic order parameter   bilinear in fermions will have $\bq$ and  $\omega$ at its arguments.   Let us now  increase the overall dimensions by   $\e$  so that we have one isospin direction,  $1+\e$ Dirac dimensions and $2+\e$ directions for $\bq$. Along with $\omega$ we have a total of $D=2+\e$ dimensions for the $(\omega,  \bk) $ vector (on which the fermion fields depend ) and $D=3+\e$ dimensions for the $(\omega, \bq)$ vector on which the order parameter depends.  The case $\e =1$ corresponds to our target, the Fermi line in $d=3$ with two Dirac dimensions.

The  action for this theory (with Fermi velocity $v_F=1$)
\beq
S_0= \int_{-\infty}^{\infty} {d\omega \over 2\pi} \int_{|\bk |<\Lambda} d^{1+\e}\bk \int d \theta \bar{\psi}(\omega \bk \theta ) \left[i \omega  - k\right] \psi (\omega \bk \theta )\label{freeaction}
\eeq
is invariant under the RG transformation:
\beqr
\Lambda&\to & s^{-1}\Lambda\\
\omega'&=&s\omega\\
\bk'&=& s \bk\\
\psi (\omega \bk \theta )&=& s^{(3+\e) /2}\psi' (\omega' \bk' \theta )
\eeqr
The angle $\theta$  that parameterizes the Fermi surface does not renormalize. Note that in an actual problem the Dirac line (Fermi surface) may not be a circle. The universal answers calculated here as well as the kinematical implications  in the limit $\La /K_F\to 0$  are insensitive to this.

Let us now add to this the BCS interaction. Recall from Refs. \cite{rg,rsrg} that in the limit  $\La / K_f \to 0$, the requirement that all four lines lie in the bandwidth and obey momentum conservation means  the only possible four-Fermi interactions are those with either nearly forward scattering described by a function $u(\theta_1 , \theta_2)$ (which will become $u(\theta_1 - \theta_2)$ for a circular Fermi surface) and those with nearly zero incoming momentum described by $v( \theta_1 , \theta_3)$ (which will become $v(\theta_1 - \theta_3) $ for a circular fermi surface). We say "nearly" in both cases because deviations of order $\La$ are kinematically allowed. These amplitude  $u$ will always be possible since if the two incoming lines lie in the bandwidth, (nearly ) forward scattering will ensure the two final lines satisfy momentum conservation and lie in the bandwidth. The same logic works for  $v$ as long as we pick one incoming and one outgoing line in the bandwidth, for their opposites  in the Cooper pair will satisfy momentum conservation and lie in the bandwidth as long as $E(-\bk ) = E(\bk )$.

By power counting the interactions $u$ and $v$ that survive the restriction to the narrow bandwidth near the Fermi surface scale as
\beq
v'= s^{-\e}v\label{uscale}
\eeq
and similarly for $u$. Thus these are irrelevant for all co-dimension $d_c > 1$, and the low energy physics is that of the free fermion theory.

Let us now study a phase transition that occurs at finite interaction strength that gaps out all the fermions. We consider a superconducting transition driven by increasing $v$.
We consider a simple model where $v$ is constant except for the antisymmetry demanded by the Pauli principle:

\beqr
S_4&=& {v\over 4}  \int \prod_{i=1}^{3}\int_{-\infty}^{\infty} {d\om_i \over 2\pi} \int_{|\bk |<\Lambda}^{} d^{1+\e}\bk_i \int d \theta_i \nonumber \\ & & \e_{\al \be} \e_{\ga \de} \bar{\psi }_\al (4)\bar{\psi }_\be (3) \psi_\ga  (2) \psi_\de (1)
\eeqr
where labels like $(2)$ are shorthand for  angles, momenta and frequency labeled $2$, and there is no integral over $(4)$ which  is  required to equal $(1+2 -3)$  by conservation laws. If $v$ is large the fermions will pair and a superconducting state will result. We now study the transition to this state to leading order in an $\e$ expansion for small $\e$.

If we eliminate the modes between $\La$ and $\La /s$ we get an all too familiar loop correction
\beq
v'= s^{-\e}v ( 1 + v \ln s )\label{uscale2}
\eeq
where the loop may be done in $D=2$ (which is $\e =0$) since we expect the $v$ in front of it  to be of order $\e$ at the fixed point.
Thus
\beq
\be (v) = {dv \over d \ln s }= -\e v + v^2
\eeq
 with a fixed point
 \beq
 v^*=\e. \label{fp}
 \eeq
 We can  now read off the correlation length exponent:
 \beq
 \nu = 1/\beta'(v^*) = 1/\e \label{nu}.
 \eeq
 We need to find the exponent that corresponds to the "magnetic field" that couples to the order parameter $ \Delta (\om , \bq )= \langle (\psi \psi \rangle )_{\om , \bq}$. So let us consider a source term
 \beqr
  S_{h}&=& \int_{|\bk |<\La } d^{1+\e} {\bk}\int d\theta  \int_{-\infty}^{\infty}{d \omega'\over 2 \pi}\nonumber \\ & & h (\omega , \bq )  \psi (\omega', \bk , \theta ) \psi (-\omega'+w, -\bk +\bq, -\theta  )
  \eeqr
  where $-\theta$ is the angle on the Fermi surface where the  momentum is opposite to that at   the point $\theta$.

  In this expression, one field $\psi$ lies inside the bandwidth and the other has the opposite frequency and momentum plus a tiny $\om , \bq $.
  It is readily verified that at tree level
  \beq
   h'= h s
   \eeq independent of $\e$.
   At one loop, the source $h$ couples to $ \psi \psi$ via a particle-particle bubble (Figure 1) whose momenta lie in the shell being eliminated. Doing the calculation at $\e =0$ due to the $v^*$ that multiplies the bubble,
   we find
   \beq
   h'= h s ( 1 + v^* \ln s ) \simeq h s^{1+\e}\label{scaleh}
   \eeq

\begin{figure}
\includegraphics[width=8cm]{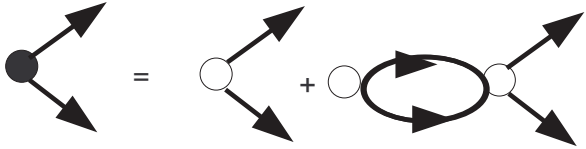}
\caption{  Renormalization of the field that couples to the BCS order parameter at one loop. Open (closed) circle is the bare (renormalized) vertex . } \label{oneloop}
\end{figure}

   All we need now is the scaling law for $E_{sing}$ the singular part of the ground state energy per degree of freedom. Recall the argument. If we eliminate some high energy variables in a path integral, it will make  an additive nonsingular contribution  the free energy, to be carried on to the next stage of mode elimination.  Since  the {\em singular} part is the same, the singular part of the energy {\em per degree of freedom} will scale as
   \beq
    E_{sing} (\delta v , h ) =s^{-D} E_{sing} (\delta v s^{\e },h's^{1+\e})
    \eeq
    where $\delta v = v -v^*$ and $D$ is the number of dimensions along which we have thinned out degrees of freedom and rescaled frequency or momentum. The answer in our case is clearly $D= 2+\e$.  Henceforth the subscript in $E_{sing}$ will be dropped. Thus we write
     \beq
    E(\delta v , h ) =s^{-2-\e} E(\delta v s^{\e },hs^{1+\e})
    \eeq
    By taking appropriate derivatives with respect to $h$ and $\delta v $   and then setting  $s= (\delta v)^{-1/\e}= (\delta v)^{-\nu}, h=0$  we find
\beqr
\alpha &=& 1 - {2 \over \e}\\
\gamma &=& 1\\
\eta &=& 2-\e\\
\beta &=& 1/\e \\
\nu &=& 1/\e
\eeqr
Some comments are in order. By $\alpha$ we mean the exponent for the second derivative of $E$ with respect to the control parameter $\delta v$ used to tune the transition. Note that $\alpha < 0$.   The exponents $\beta$ and $\gamma$ come from the first and second derivatives with respect to $h$. One can find $\eta$, the anomalous exponent for the order parameter correlation  by directly computing it to one loop at $v^*=\e$ or by using the  relation
\beq
\gamma = (2 -\eta) \nu. \label{hyper}
\eeq

We emphasize that the fermionic RG gives a canonical  description of the critical point. In particular
conventional hyperscaling equations are satisfied with space-time dimensionality $d_c + 1$.

Let us ask what we would find had we chosen the Hertz-Millis  route of integrating out the fermions to get an action for the pairing field $\Delta$. We would find
\beqr
S_{\Delta}&=& \int {\Delta}^* (\omega , \bq )((q^2 + \omega^2)^{\e/2}+ r)\Delta ( \omega, \bq ) {d \omega\over 2 \pi} d^{2+\e}\bq \nonumber \\ & & + u \Delta^* \Delta \Delta^* \Delta\mbox {\  terms+ ...}\label{bosonaction}\eeqr
Note that $\bq $ lives in $2+\e$ dimensions and we have a
$D=3+\e$ theory. It has to reproduce a fixed point that we saw is of effective dimensionality $D= 2+\e$. Let us see how far we can go with the quadratic part of the action which is just the Cooper pair susceptibility.

First, the tree level scaling for the critical $r=0$ theory is
\beq
\Delta  = s^{(3+ 2\e)/2}\Delta' \ \ \  \mbox{which means  } \  h' = s^{(3+ 2\e)/2}h\label{scaleh2}
\eeq
from which it follows that
\beq
r'= rs^\e\ \ \ \ \    \mbox{which means  } \  \nu = { 1\over \e}
\eeq
in agrement with the old answer, even though $\Delta$ scales  differently now.

From Eq. \ref{scaleh2} it follows that the  quartic coupling $u$ in Eqn. \ref{bosonaction} is   highly irrelevant {\em if  it is  nonsingular in its $\om$ and $\bq$ dependence. }   But it is   singular due to the fermion loop that generates it, but  still irrelevant. Let us now calculate $\beta$ paying attention to this dangerously irrelevant term.

If we evaluate the fermionic loop contribution to $u$, with all external $\bq$ and $\om$ equal to zero we will find (from a box diagram of 2 pairs of fermions of equal  and opposite $(\om , \bk)$),
\beq
u \simeq \int {d \om d^{1+\e}\bk \over (\om^2 + k^2)^2}
\eeq
The infrared divergence of this diagram tells us  that \beq
u (\om , \bq) \simeq {u_0 \over (\sqrt{\om^2+  q^2} )^{2-\e} }.\eeq
 From this and Eqn. \ref{bosonaction} we may deduce that
\beq
u_{0}(s) = s^{-1}u_0.
\eeq

As in the case of $\phi^4$ theory above $4$ dimensions the $u$ will be dangerously irrelevant and it will be necessary to keep it to obtain correct results for some exponents.

Turn now to the energy density which scales as follows
\beq
E(r,h,u_0)= s^{-(3+\e)}\ E(r s^\e, hs^{({3 \over 2}+ \e)},u_0s^{-1})\label{hyperfake}
\eeq
The exponents $\nu,\gamma$ and $\eta$ can be deduced by working in the non-superconducting state ({\em i.e} $r > 0$). Then we may safely set $u =0$, and the results are identical to that obtained above by the fermionic RG.

The exponents $\beta$ and $\alpha$ require retaining the irrelevant interactions.
From the first derivative of $E$ with respect to $h$ we obtain the order parameter $m= <\Delta>$:
\beq
m(r, 0, u_0)=s^{-(3+\e)}s^{({3 \over 2}+ \e)} m(r s^\e, 0 ,u_0s^{-1})\label{betam}
\eeq
and the mean-field result far from the transition when $rs^\e$ is of order unity
\beq
m(r s^\e, u_0s^{-1})\simeq \sqrt{{r s^\e \over u_0s^{-1}}}
\eeq
Setting $rs^\e=1$ we find
\beq
\beta =1/\e
\eeq
again in agreement with the fermionic RG. Finally  $\alpha$ (exponent for the second derivative with respect to the control parameter) may  be obtained {\em within mean field theory} by examining the ground state energy in the presence of a uniform condensate $\Delta$ which assumes the {\em singular} form
 \beq
 E= -r |\Delta |^2 + u |\Delta |^{2+\e}
 \eeq
 in terms of $r$ and $u$ which are viewed as fixed (and not running) parameters. At the minimum $\Delta \simeq (r/u)^{1/\e}$ (so that $\beta = 1/\e$) while
 the second  $r$ -derivative of the minimum energy gives $\alpha = 1- {2 \over \e}$,   in agreement  with the fermionic RG.  Note that the Gaussian action for $\Delta$ in the disordered phase describes the fluctuation correction which is subdominant to the mean field answer.

The agreement of the fermionic and bosonic versions is  remarkable and seems unlikely  given that they describe theories in different number of effective dimensions ($2+\e$ for fermions and $3+\e$ for bosons). Despite this and the fact that $\Delta$ scales differently in the two versions, all exponents are the same {\em order $\e$}. It is always possible that something differs at higher orders in the  $\e$ expansion.

There are reasons for believing that the one-loop order $\e$ results are exact. As stressed in
Ref. \cite{rsrg} a narrow bandwidth problem has a small parameter $\La /K_F$ that plays the role of $1/N$. It is well known that in the large $N$ limit the one-loop beta function is exact.
Another way to say this is that one may simply do the RPA.  As explained in Ref. \cite{rsrg} the RPA sum over repeated bubbles
becomes exact because there is no phase space for any other diagram in the limit $\La /K_F \to 0$.

Despite this agreement we emphasize that the fermionic approach is much more {\em natural} as it properly gives a
scale invariant fixed point with a correct effective dimensionality of $d_c + 1$. In particular all hyperscaling equalities are satisfied with this effective dimension. The Hertz-Millis theory does not, at first sight, look like a theory that will satisfy any such scaling. However after including for the complications of the dangerously irrelevant interactions this hidden effective dimensionality is correctly captured. Indeed the Hertz-Millis description needed to be more complicated in order for such a $d+ 1$-dimensional theory to scale as though it lived in a different lower dimension $d_c + 1$.

Though we will not discuss it here clearly we can now go ahead and compute universal physics associated with the transition at finite temperatures to study the quantum critical region\cite{qc}. One aspect of this theory is worth
considering. What is the fate of the fermions themselves right at the critical point? This may be straightforwardly studied by considering the self-energy of the fermions at the fixed point. We find that the fermionic quasiparticles
are not destroyed even at this non-trivial interacting fixed point. In other words the quasiparticle residue $Z$ stays non-zero all the way upto and including the critical point. The disappearance of the generalized Fermi surface on entering the superconducting state thus happens without ever destroying the fermionic quasiparticle. This is a different route to killing a Fermi surface from the Mott-like transitions considered in Ref. \onlinecite{critfs} where $Z$ vanishes at the critical point\cite{BrRice}. Note that the Fermi surface stays sharp at the present critical point as expected on general grounds\cite{critfs}.

To conclude we studied fermion systems with generalized `Fermi surfaces' with co-dimension $d_c$ in $d$ space dimension in the presence of interactions. Not surprisingly weak short range interactions were irrelevant for all $d_c > 1$. We studied a non-trivial superconducting transition out of such a phase through a controlled co-dimension expansion using the fermionic RG technique. The resulting theory scales with effective dimensionality $d_c + 1$. We also demonstrated that this result can be recovered within the Hertz-Millis approach despite the different aparent dimensionality.
More generally studying phase transitions of Fermi surfaces with general co-dimension $d_c$ in $d$ space dimensions is a potentially useful theoretical device. In the future it might be interesting to understand the Stoner transition or problems with gauge fields by generalizing to $d_c > 1$. Perhaps such generalizations will lead to controlled expansion methods
suitable for such problems.

We are grateful to Ganpathy Murthy for very useful conversations and the NSF for grants DMR-0705255 (TS), and DMR-
0103639 (RS). RS is grateful to the Center for Condensed Matter Theory  at MIT for  hosting his sabbatical  in Spring 2008, which made this work possible.

 \end{document}